\documentclass[referee]{aa} 

\usepackage{color}
\usepackage{graphicx}
\usepackage{natbib}
\usepackage{txfonts}

\newcommand{\wavenb}{cm$^{-1}$}

\newcommand{\myemail}{nicolas.iro@obspm.fr}
\newcommand{\HD}{HD209458}%

\begin{document}
   \title{A Time-Dependent Radiative Model of HD209458b}

   \author{N. Iro, B. B\'ezard
          \inst{1}
          \and
          T. Guillot\inst{2}
          }
   \offprints{N. Iro, \\ \email{\myemail}}

   \institute{LESIA, Observatoire de Paris-Meudon, place Jules Janssen, 92395 Meudon Cedex, FRANCE
         \and
             Observatoire de la C\^ote d'Azur, CNRS UMR 6202, BP4229,
             06304 Nice Cedex 4, FRANCE\\
             }

   \date{Received ; accepted }

   \abstract{ We present a time-dependent radiative model of the
atmosphere of HD209458b and investigate its thermal structure and
chemical composition.  
In a first step, the stellar heating profile and radiative
timescales were calculated under planet-averaged insolation
conditions. We find that 99.99\% of the incoming stellar flux has been
absorbed before reaching the 7 bar level. Stellar photons cannot
therefore penetrate deeply enough to explain the large radius of the
planet. We derive a radiative time constant which
increases with depth and reaches about 8 hr at 0.1 bar and 2.3 days at 1 bar.
Time-dependent
temperature profiles were also calculated, 
in the limit of a zonal wind that is independent on height (i.e. solid-body rotation)
and constant absorption coefficients.
We predict day-night variations of the effective temperature of
$\sim$600 K, for an equatorial rotation rate of 1 km s$^{-1}$, in
good agreement with the predictions by
\cite{ShowGuillot}. This rotation rate yields day-to-night temperature
variations in excess of 600 K above the 0.1-bar level. These
variations rapidly decrease with depth below the 1-bar level and
become negligible below the $\sim$5--bar level for rotation rates of
at least 0.5 km s$^{-1}$.  At high altitudes (mbar pressures or less),
the night temperatures are low enough to allow sodium to condense into
Na$_2$S.  Synthetic transit spectra of the visible Na doublet show a
much weaker sodium absorption on the morning limb than on the evening
limb. The calculated dimming of the sodium feature during planetary
transites agrees with the value reported by \cite{Charbonneau}.

   }

   \maketitle
%

\section{Introduction}

The discovery of \HD b (\citet{Charbonneau00}; \citet{Henry00}) allows
for the first time probing directly the structure of a planet outside
our Solar System. Indeed, the fact that it transits in front of its
star allows both the measurement of its radius and the spectroscopic
observation of its atmosphere. 

Quantitatively, \HD\ is a G0 subgiant, with a mass $M_\star=1.06
M_\odot$, radius $R_\star=1.18 R_\odot$ and age $t_\star=5.2$\,Gyr,
with uncertainties of 10\% or more \citep{CodySass02}. The planet
orbits the star in 3.524739 days \citep{Rob_Arenou}, at a
distance $\sim 0.045$\,AU. Models of the star imply that its radius is
$R$~= 92\,200--109\,000~km (about 40\% more than Jupiter) for a mass
$M=0.69\pm 0.02\,M_{\rm J}$ \citep{Brown_al01,CodySass02}.

The relatively large radius of the planet appears difficult to explain
using standard evolution models \citep{Boden01,Boden03,GuillotShow,Baraffe}.
\citet{GuillotShow} concluded that, for a realistic model of the outer 
atmosphere irradiated by the parent star, an extra energy source is needed
at deep levels to explain \HD 's radius. It could result from the penetration
of a small fraction of the stellar flux at pressures of tens of bars or from
kinetic energy produced from the stellar heating and transported 
downwards to the interior region. The same conclusion was reached by 
\citet {Baraffe} from model calculations directly coupling the irradiated
outer atmosphere and the interior. However, the possibility that
systematic errors both in the determination of the stellar radius and
atmospheric properties of the planet remains \citep{Burrows03}. 

In order to accurately model the evolution of extrasolar giant
planets, and hence gain information on their composition, one has to
understand how their atmospheres intercept and reemit the
stellar irradiation. The problem is especially accute for planets like
51 Peg b and \HD b (hereafter Pegasides) which are believed to be
locked into synchronous rotation \citep{Guillot96} due to their
proximity to their star. As a consequence the amount of irradiation
received on the day side and the insolation pattern are unlike what 
is experienced by any planet in the Solar System. 

Several studies have investigated the radiative equilibrium structure 
of Pegasides, treating the atmosphere as a one--dimensional column receiving 
an average flux from above and a smaller intrinsic flux from below. 
In these models, the stellar heat is either evenly distributed over the 
entire planet \citep{Seag_Sass98,Seag_Sass00,Gouken}, or redistributed 
only over the day side \citep{Sudarsky03,Baraffe,Burrows03}, or even not 
redistributed at all \citep{Barman01,Burrows03}. In fact, in a planetary 
atmosphere, winds carry part of the stellar heat from the day side 
to the night side and from equator to poles, so that the thermal 
structure depends not only on the insolation pattern but also on the
dynamics.

Two recent dynamical studies \citep{ShowGuillot,Cho} have shown 
that the atmospheric structure probably presents strong latitudinal 
and longitudinal variations in temperature (and hence composition).
These two investigations differ in several respects:
\cite{ShowGuillot} used analytical arguments to show that the
atmospheres of Pegasides should have strong winds ($\sim
1\rm\,km\,s^{-1}$) and relatively strong day-night and 
equator-to-pole temperature contrasts ($\sim 500\,$K near optical depth
unity). These estimates are confirmed by preliminary 3D simulations
done using the EPIC (Explicit Planetary Isentropic Coordinate) model 
with a radiative time constant of 2.3 days. These simulations tend 
to yield a prograde equatorial jet and thus imply that the atmosphere 
could superrotate as a whole (as is the case for Venus). 
On the other hand, \cite{Cho} solved 2D shallow-water
equations, assuming a
characteristic wind speed, related to the mean kinetic energy, of 50
to 1000\,m\,s$^{-1}$
and a radiative equilibrium time of 10 days.
Their simulations yield a circulation which is
characterized by moving polar vortexes around the poles and $\sim$ 3
broad zonal jets.  For the highest speed considered (1 km s$^{-1}$),
the temperature minimum is about 800 K and the maximum temperature
contrast is $\sim 1000$\,K.

The goal of the present study is to apply a time-dependent radiative
transfer model to the case of \HD b in order to: (i) determine its mean
temperature structure and stellar heating profile; (ii) estimate 
the characteristic radiative heating/cooling timescale as a function 
of depth; (iii) model longitudinal temperature variations assuming a 
solid rotation; (iv) infer consequences for the variations 
in chemical composition in light of the spectroscopic transit observations.

In Section~\ref{model}, we present our time-dependent radiative
transfer model. We then apply the model to \HD b. In
Section~\ref{sec:radeq} we compare models obtained for averaged
insolation conditions, and calculate the corresponding radiative
timescales. In Section~\ref{sec:timedep}, we calculate the
longitude-dependent thermal structure of the atmosphere by assuming a solid
body rotation, mimicking a uniform zonal wind. The variations in the chemical
composition induced by temperature variations are investigated with
the emphasis on the condensation of sodium.  A summary of the results
and a conclusion are presented in Section~\ref{conclusions}.


\section{The Atmospheric Model}
\label{model}

%

%

\subsection{Physical problem}

In a one-dimensional model, the evolution of the temperature profile in 
an atmosphere under hydrostatic equilibrium is related to the net 
flux $F$ by the following energy equation:

\begin{equation}\label{eq:dTdt1}
\frac{d T}{d t}=\frac{m g}{C_{\rm p}}\left(\frac{d F} {d p}\right) \quad {\rm,}
\end{equation}
where $m g$ is the mean molecular weight and $C_{\rm p}$ is the mean 
specific heat. 
The net flux is divided into the thermal flux emitted by the atmosphere 
$F_{\rm IR}$ (upward - downward) and the net stellar flux $F_\star$ 
(downward - upward), so that $F=F_{\rm IR} - F_\star$. Equation~\ref{eq:dTdt1}
can then be rewritten as:

\begin{equation}\label{eq:dTdt}
\frac{d T}{d t}=\left(h(p) - c(p)\right) \quad {\rm,}
\end{equation}
where $h(p) = -\frac{m g}{C_{\rm p}}\frac{d F_\star}{d p}$
is the heating rate and
$c(p) = -\frac{m g}{C_{\rm p}}\frac{d F_{\rm IR}}{d p}$ is the cooling rate.

Radiative equilibrum corresponds to a steady-state solution of the energy
equation and is thus obtained by setting the left-hand term of Eqs.~1--2 
to zero. In this case, the flux is conservative and heating and cooling
rates are equal at any level in the atmosphere. In a first step, we 
calculate the radiative equilibrium solution for \HD b, using a
planetary-averaged stellar irradiation. In a second step, we 
investigate the variations in the temperature profile due to 
the time-varying insolation, which requires solving Eq.~\ref{eq:dTdt}.

\subsection{Numerical method}

We use the atmosphere code described in \citet{Gouken}, with, in 
the present model, an atmospheric grid of $N=96$ levels from 
$3\times 10^3$ to $1\times10^{-6}$ bar.  To calculate $F_\star$ as 
a function of pressure level $p$, we solve the radiative equation 
of transfer with scattering in the two-stream approximation and 
in plane-parallel geometry using a monochromatic line-by-line code. 
The boundary condition is that the incident downward flux 
$F_\star^\downarrow(0)$ is given by: 
\begin{equation}\label{eq:solarflux}
F_\star^\downarrow(0)=\alpha \pi 
\left(\frac{R_\star}{a}\right)^2B_\nu(T_\star) \quad {\rm,}
\end{equation}
where $R_\star$ is the star's radius, $a$ the distance of the planet
to the star's surface, and $B_\nu(T_\star)$ is the monochromatic Planck
function at the star's brightness temperature $T_\star$.  To solve for 
steady-state radiative equilibrium, we consider disk-averaged 
insolation conditions, and thus use $\alpha = \frac{1}{4}$. 
For our time-dependent calculations, we use
$\alpha$~=~$\pi$~max[$\cos(\lambda)$,0], where $\lambda$ is the 
zenith angle of the star. The stellar flux is calculated 
from 0.3 to 6 $\mu$m (1700-32000 \wavenb ).  
It is assumed that shortward of 0.3 $\mu$m photons are either 
scattered conservatively or absorbed above the atmospheric grid 
($p<1$ $\mu$bar) and thus do not participate to the energy budget.

The planetary thermal flux is calculated from 0.7 to 9 $\mu$m.  At
level $p_{\rm i}$, it can be expressed as:

\begin{equation}
\label{eq:Fth}
F_{\rm IR}= \sum_k \sum_{j=1}^{N} a_{\rm i,j,k} B_{\rm k} 
\left(T_{\rm j}\right) \quad{\rm ,}
\end{equation}
where $ B_{\rm k} \left(T_{\rm j}\right)$ is the Planck function at the
temperature $T_{\rm j}$ of pressure level $p_{\rm j}$ and at frequency 
$\nu_{\rm k}$, and $a_{\rm i,j,k}$ (= $a_{\rm j,i,k}$) is a coupling 
term related to the transmittance between levels $p_{\rm i}$ and $p_{\rm j}$, 
averaged over a frequency interval centered at $\nu_{\rm k}$ \citep{Gouken}.  
We use an interval width of 20 \wavenb .  These coefficients are calculated
through a line-by-line radiative transfer code with no scattering.
Below the lower boundary located at level $p_1$, an isothermal layer
of infinite optical depth is assumed with a temperature $T_{\rm gr}$.
It does not represent the actual temperature
profile in the interior. In fact, for numerical reasons, what is exactly set in the model
is the upward flux at the lower boundary.

\subsubsection{Steady-state case}

Starting from an initial guess temperature profile and the associated
abundance gas profiles, stellar and thermal fluxes are computed on the
atmospheric grid.  The heating and cooling rates, $h(p_{\rm i})$ and
$c(p_{\rm i})$, defined above are then calculated.  At each level 
of the grid, the temperature is modified by an amount:
\begin{equation}
\Delta T_{\rm i} = \epsilon (p_{\rm i})\left[ h(p_{\rm i})-c(p_{\rm i})\right]
+ \eta(p_{\rm i})\left(T_{\rm i-1}-2T_{\rm i}+T_{\rm i+1}\right)
\end{equation}
The coefficient $\epsilon(p)$, analogous to a time step, increases with
pressure to account for the increase of the radiative time constant
with depth and thus to speed up convergence.  
Its value is about one hour in the upper layers and is as long as few days
in the deepest layers.

The second term in the
equation introduces some numerical dissipation needed to avoid
instabilities at high pressure levels where the atmospheric layers are
optically thick.  The coefficient $\eta (p)$ is set to $1\times
10^{-3}$ above the 0.3--bar pressure level and reaches 2\% at the lower
boundary ($p_1$).  The temperature $T_{gr}$ beneath pressure level
$p_1$ is set to $T_1+\delta T_1$, where $\delta T_1$ is calculated to
yield $F_{\rm IR}(p_1)=F_{\rm int}$.  The thermal flux is then
calculated using the updated temperature profile and the procedure is
iterated until the condition $h(p_i)=c(p_i)$ is fulfilled at each
level with a precision of 5\%.

A convective adjustment is then applied to the solution profile
in regions where the
radiative gradient exceeds the adiabatic value.  In these convective
regions, the gradient is set to the adiabatic value and the energy
flux then includes both convective and radiative components (
Eq.~1 does not hold anymore).

After convergence, the equilibrium gas profiles are re-calculated
using the solution temperature profile.  The stellar flux deposition
and the $a_{\rm i,j,k}$ coefficients involved in the calculation of the
thermal flux are then re-computed and a new temperature profile is
obtained from the iterative process described above.  If this new
solution profile is close enough to the previous one (with a precision
of 10~K), we retain it as the radiative equilibrium solution.  If not,
the whole iterative process is continued.

\subsubsection{Time-dependent case}

To investigate the atmospheric response to a time-varying insolation, we
set the incident stellar flux according to Eq.~\ref{eq:solarflux} with 
$\alpha$~=~$\pi$~max[$\cos(\lambda)$,0] and $\lambda$~= 
$\frac{2 \pi t}{T}$, $T$ being the rotation period of the atmosphere. 
This insolation pattern represents the variation of the stellar flux during the
day, averaged over latitude (i.e., along a meridian). It yields the 
same daily insolation as the steady-state planet-average case 
($< \alpha >_{\rm t}~= \frac{1}{4}$). We then solve Eq.~\ref{eq:dTdt} 
using a time-marching algorithm with a timestep of 300 sec, 
starting from the radiative-equilibrium solution profile. Integration is
performed over several rotation periods until the temperature reaches a 
periodic state at each level.

\subsection{Opacities}

We include the following opacity sources: Rayleigh scattering (for
the stellar flux component only), collision-induced absorption from
H$_2$-H$_2$ and H$_2$-He pairs, H$^-$ bound-free, H$_2^-$ free-free
absorption, molecular rovibrational bands from H$_2$O, CO, CH$_4$, and
TiO, and resonance lines from Na and K.  Spectroscopic data used for
H$_2$-H$_2$, H$_2$-He, H$_2$O, CO, and CH$_4$ are described in
\citet{Gouken}.  An improvement over Goukenleuque {\it et al.}'s
modeling was to add absorption from the neutral alkali metals Na and K
whose important role was established by \citet{Burrows_alc}.  For the
resonance lines at 589 nm (Na) and 770 nm (K), we followed the general
prescription of \citet{Burrows_alc} to calculate their collisionally
broadened lineshape.  In the impact region, within $\Delta\sigma$ of
line center, a Lorentzian profile is used with a halfwidth calculated
from the simple impact theory: $\gamma$ = 0.071($T$/2000)$^{-0.7}$
cm$^{-1}$ atm$^{-1}$ for Na and $\gamma$ = 0.14 ($T$/2000)$^{-0.7}$
cm$^{-1}$ atm$^{-1}$ for K.

Beyond the transition detuning $\Delta \sigma$, we use a lineshape
varying as $(\nu-\nu_0)^{-3/2}$ as predicted from the statistical
theory.  We multiply this lineshape by $e^{-h(\nu-\nu_0)/kT}$ to
account for the exponential cutoff of the profile at large frequencies
($\sim$1000--3000 cm$^{-1}$ at 2000 K).  We adopt the detuning
frequencies $\Delta\sigma$ calculated by \citet{Burrows_alc}, 30
cm$^{-1}$($T$/500 K)$^{0.6}$ for Na and 20 cm$^{-1}$($T$/500
K)$^{0.6}$ for K.  H$^-$ bound-free and H$_2^-$ free-free absorption
were modeled as in \citet{Guillot94} using the prescription of
\citet{John88} for H$^-$ and \citet{Bell80} for H$_2^-$. 

For TiO, we
use absorption coefficients from \citet{Brett90}.
This
opacity data base is known to have limitations, especially for
effective temperatures $\sim 2500\,$K \citep{Allard00}.
This should have a minor impact on the structure of HD209458b
because of its relatively low effective temperature, and appears
justified in regard of the other sources of uncertainties of the
problem.

A relatively simplified chemical equilibrium is calculated using the
ATLAS code \citep{Kurucz} including the condensation of Na (as
Na$_2$S), K (as K$_2$S), Fe (as Fe), Mg and Si (as MgSiO$_3$), and assuming
that Al, Ca, Ti and V also condense with similar partial pressures as
MgSiO$_3$. The condensation curves are taken from
\citet{Fegley} and \citet{Lodders}.


An important uncertainty in any attempt to calculate the atmospheric
structure of Pegasides concerns the presence and structure (in terms
of composition and grain sizes) of silicate clouds in the atmosphere
(e.g. \citealp{Fortney03}). As discussed by \citet{ShowGuillot},
such a cloud could either occur on the day side or on the night side
depending on whether the global atmospheric circulation yields a
strong vertical advection at the substellar point or a mainly
superrotating atmosphere. Given this unknown and the very uncertain
microphysics underlying the cloud structure, we choose not no include
any additional opacity due to condensed particules in the present
model.

\subsection{Input parameters}

We assume a solar abundance of the elements \citep{Anders}.
The acceleration
of gravity was set to $g$ = 9.7 m s$^{-2}$ corresponding to a
radius of 1.35 Jupiter radius and a mass of 0.7 time that of Jupiter
\citep{Mazeh00}. To calculate the incoming stellar flux, we adopt
$R_\star$~= 1.18 $R_\odot$, $a=0.047$ AU \citep{Mazeh00}, and use the
brightness temperature spectrum of the Sun given in \citet{PierceAllen}.  

Our lower boundary condition is fixed by the intrinsic flux of the
planet $F_{\rm int} = \sigma T_{\rm int}^4$.  Standard evolution models, in
which the stellar flux is totally absorbed at low pressure levels ($<$~10 
bars), indicate an intrinsic effective temperature $T_{\rm int}$ of
approximatively 100~K at the age of \HD b.  On the other hand, 
to reproduce the relatively large observed radius of the planet, an 
intrinsic temperature as high as $\sim$300~K is needed 
\citep{GuillotShow,Baraffe}, which requires an extra source 
of energy in the interior.  In our model, we have thus considered 
two possible boundary conditions corresponding to 
$T_{\rm int}$ = 100~K (cold case) and 300~K (nominal case).

\section{Radiative equilibrium solutions}
\label{sec:radeq}

\subsection{Temperature profiles}

\begin{figure}[tb]  
\begin{center}
   \includegraphics[width=\columnwidth]{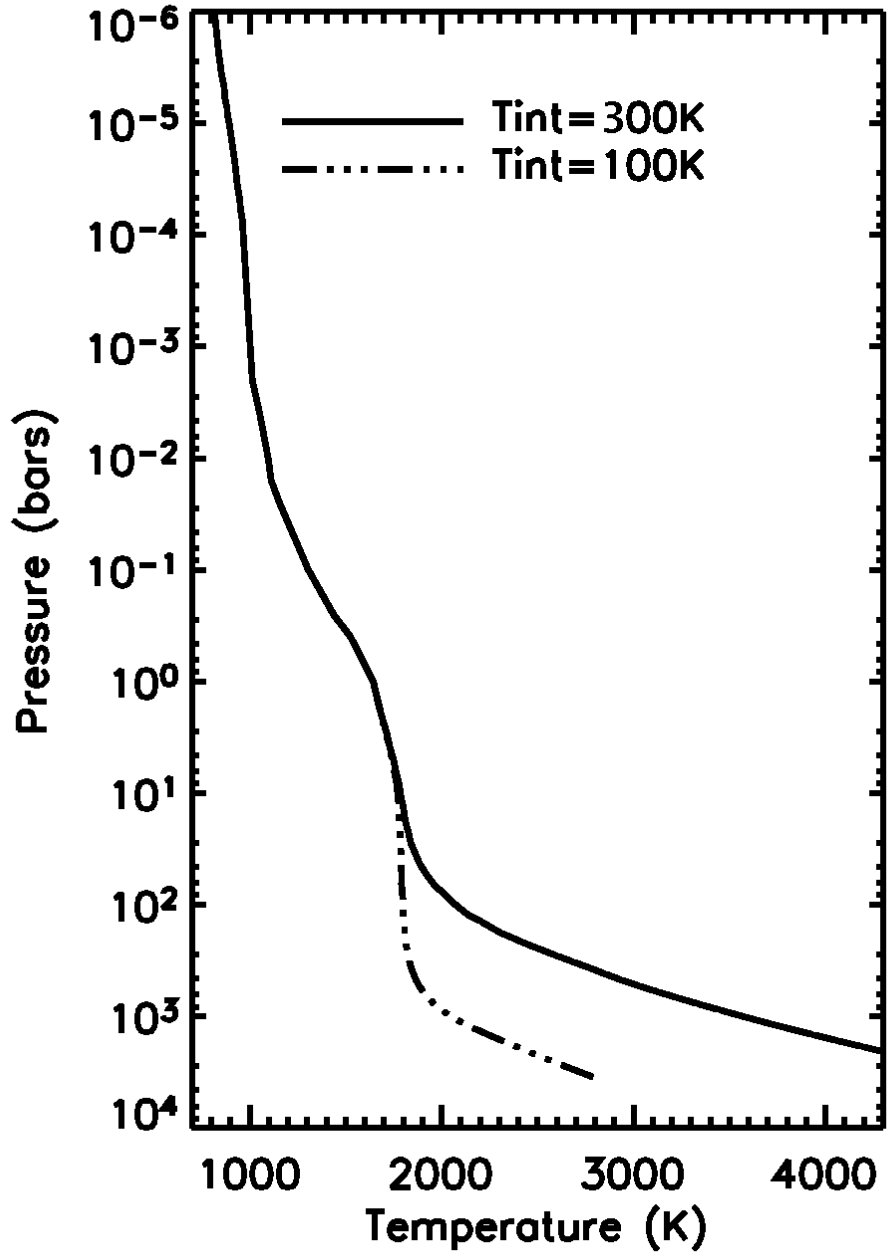}
\caption{Temperature profiles of the static disk-averaged solutions
for the two boundary conditions.\label{fig:profs} There is no
difference above the 10--bar level as the intrinsic heat flux becomes
negligible with respect to the net stellar flux.}
\end{center}
\end{figure}

We compute two atmospheric temperature profiles corresponding to
the two boundary conditions ($T_{\rm int}$ = 100~K and $T_{\rm int}$ = 300~K), 
as shown in Fig.~\ref{fig:profs}.  Above the 10--bar level, the boundary
condition does not make any difference because the incoming stellar
flux is dominant with respect to the intrinsic heat flux.  A
convective zone appears below the 1--kbar level for the cold case 
and the 0.1--kbar level for the nominal case.

We infer an effective temperature\footnote{defined from the total
flux emitted by the planet by: $F_{\rm IR} = \sigma T_{\rm eff}^4$
}
$T_{\rm eff}$ = 1340 and 1350~K
for the two boundary conditions (cold and nominal respectively), 
and a planetary Bond albedo of 0.10.  While this albedo closely agrees 
with that inferred by \citet{Baraffe}, our $T_{\rm eff}$ is lower 
because these authors redistribute the stellar flux only over the 
dayside and thus have a stellar heating twice larger than ours. 
Below the 3-mbar level, our profile is $\sim$ 600 K cooler than 
that produced by \citet{Sudarsky03} (as shown in their Fig.~26). 
This difference partly results from the twice larger insolation 
used by these authors who chose the same redistribution of stellar 
heat as \citet{Baraffe}. However it can only produce 15-20\% larger
temperatures as illustrated by \citet{Sudarsky03}'s comparison of 
profiles with different incident flux weighting (shown in their 
Fig.~28). The expected temperature increase below the 3-mbar level is 
thus limited to 150--300~K. On the other hand, we note that these authors
used an intrinsic temperature of the planet of 500~K, hotter than ours, 
which can contribute to the difference with our model at least below the
$\sim$0.5--bar level. Above the 0.3--mbar level, the difference is only
$\sim$150~K and can be accounted for by the factor of two in the 
stellar flux. Another difference between the two models is the incorporation 
of silicate and iron clouds in \citet{Sudarsky03}'s model whereas ours is 
cloud--free. These clouds, with bases at 5--10 mbar in their model, result
in a cooler atmosphere below the $\sim$30--mbar level and a hotter 
atmosphere above, and therefore are not the source of the discrepancy 
in the lower atmosphere. The (cloud--free) profile calculated by 
\citet{Baraffe} is quasi--isothermal at $\sim$1800~K between 10$^{-4}$
and 100 bar whereas ours increases from 1000 to 1800~K for the same
$T_{\rm int}$=100~K. Lowering the insolation by half in \citet{Baraffe}'s
model should bring it down to $\sim$1500~K but cannot produce 
the gradient we have or the even steeper one of \citet{Sudarsky03}.
Below the 0.3--bar region, our profile is intermediate between those
of \citet{Sudarsky03} and  \citet{Baraffe} after correcting to first
order for their twice larger stellar heating.

The preferred boundary condition used by \citet{GuillotShow}
for \HD b's radius is: $T(p=3 {\rm bar})
=T_{\rm isolated}(T_{\rm eff},g)-1000\textrm{ K}$, where 
$T_{\rm isolated}(T_{\rm eff},g)$ is the temperature of an 
isolated planet of same effective temperature $T_{\rm eff}$
and gravity $g$ as calculated by \citet{Burrows97}.
This is relatively close to our solution profile,
which implies a 3--bar temperature that is 1200~K less than 
in the isolated case.

\subsection{Penetration of the stellar flux and spectra}

\begin{figure}[tb]  
\begin{center}
   \includegraphics[width=\columnwidth]{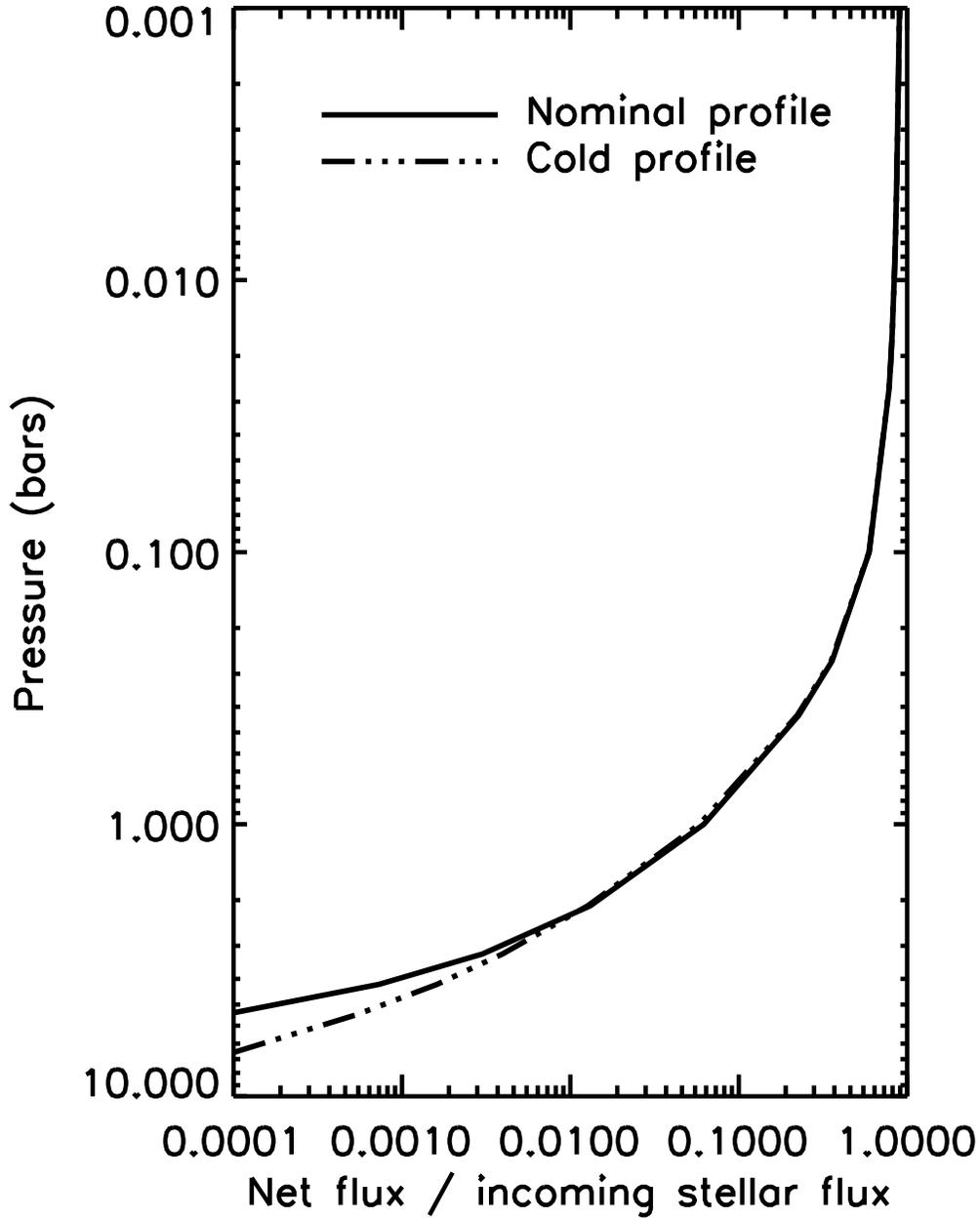}
\caption{\label{fig:penet}Net stellar flux as a function of pressure level for the Cold and Nominal profiles.}
\end{center}
\end{figure}

Figure~\ref{fig:penet} shows the net stellar flux as a function of
pressure level for the ``cold'' and ``nominal'' profiles. We find that
90\% of it is absorbed at the $\sim$0.9-bar level, 99\% around the
2-bar level, and 99.99\% at the $\sim$7-bar (resp.\ 5--bar) level in
the cold (resp.\ nominal) case. 
At this level and deeper, the stellar flux becomes smaller than the
intrinsic flux (calculated in the cold case) and cannot affect
the temperature profile in a significant way.
At deep
levels, the stellar flux decreases more rapidly with depth in the
nominal case than in the cold case due to the larger abundances of
H$_2^-$ and H$^-$ ions and of TiO. \citet{GuillotShow} estimated that
a penetration of 1\% of the stellar flux to $p~\sim$~100~bar in the
``cold'' case would allow the radius of \HD b to be explained without
any other energy dissipation. Clearly our calculations indicate that
this is not the case and that the large atmospheric opacity beneath
pressure levels of a few bars prevents any significant fraction of the
stellar heat from reaching directly the $\sim$100~bar
region. Therefore, provided that the error bars on the planet's radius
are not underestimated, we confirm the need for an additional heat
source at deep levels as advocated by \citet{Boden01}, \citet{GuillotShow} and
\citet{Baraffe}.

\begin{figure}[tb]  
\begin{center}
   \includegraphics[width=\columnwidth]{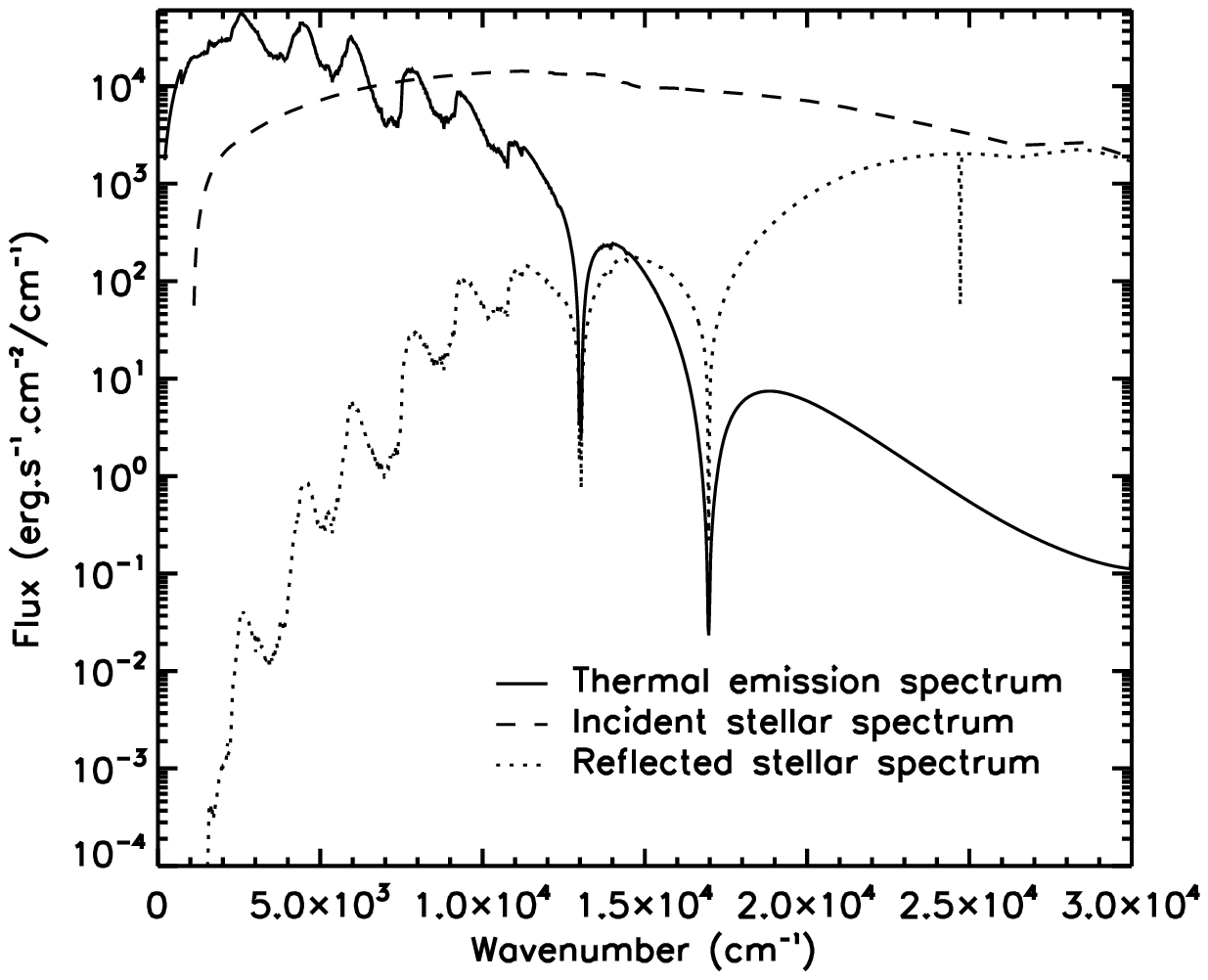}
\caption{Reflected and thermal emission spectra of the static
disk-averaged solution for the nominal boundary
condition, in logarithmic scale\label{fig:flux}.  The incident stellar spectrum is shown for
comparison.}
\end{center}
\end{figure}

The reflected and thermal spectra of the planet are shown in
Fig.~\ref{fig:flux}.  The thermal spectrum is dominated by the water
vapor bands, but CO absorption bands around 2100, 4300, and 6400 
\wavenb\ are visible. It exhibits spectral windows centered at 
2600 \wavenb\ (3.8 $\mu$m), 4500 \wavenb\ (2.2 $\mu$m), 6000
\wavenb\ (1.7 $\mu$m), 7800 \wavenb\ (1.28 $\mu$m), and 9400 \wavenb\ 
(1.07 $\mu$m). The most intense one at 3.8 $\mu$m is limited on the
short-frequency side by the (1--0) CO band and on the high-frequency side
by the $\nu_3$ H$_2$O band. The stellar reflected spectrum peaks around 
28\,500 \wavenb\ (0.35 $\mu$m). At wavenumbers less than $\sim$18\,000
\wavenb , it is almost completely absorbed due to weaker Rayleigh scattering
and strong atmospheric absorption. In particular, in the region of 
the alkali lines (13\,000--17\,000 cm$^{-1}$), the flux emitted by
the planet is at minimum while that of the star is large. This 
emphasizes the importance of alkali line absorption in the 
energy balance of the Pegasides, as first recognized by \citet{Burrows_alc}.\\

\subsection{Radiative Timescales}
\label{trad}

To characterize the radiative time constant $\tau_{\rm rad}$ at 
a given pressure level $p_0$, we applied a gaussian perturbation 
to the radiative equilibrium temperature profile and let it 
relax to its equilibrium state using Eq.~\ref{eq:dTdt}. The 
perturbation has the form $\Delta T(p) = \Delta T_0~2^{-\left[2 \ln \frac{p}{p_0}\right]^2}$,
i.e.\ a full width at half maximum of one scale height. 
The radiative time constant is calculated from the Newtonian cooling
equation:

\begin{equation}
\frac{\Delta T}{\tau_{\rm rad}}=-\frac{\partial (\Delta T)}{\partial t} 
\quad ,\end{equation}
where $\Delta T$ is the temperature deviation from the equilibrium
profile at level $p_0$.  We verified that the Newtonian cooling
approximation is justified and yields the same $\tau_{\rm rad}$ as far as 
$\Delta T_0$ is small (typically less than 5\% of the equilibrium temperature).

\begin{figure}[tb]   
\begin{center}
   \includegraphics[width=\columnwidth]{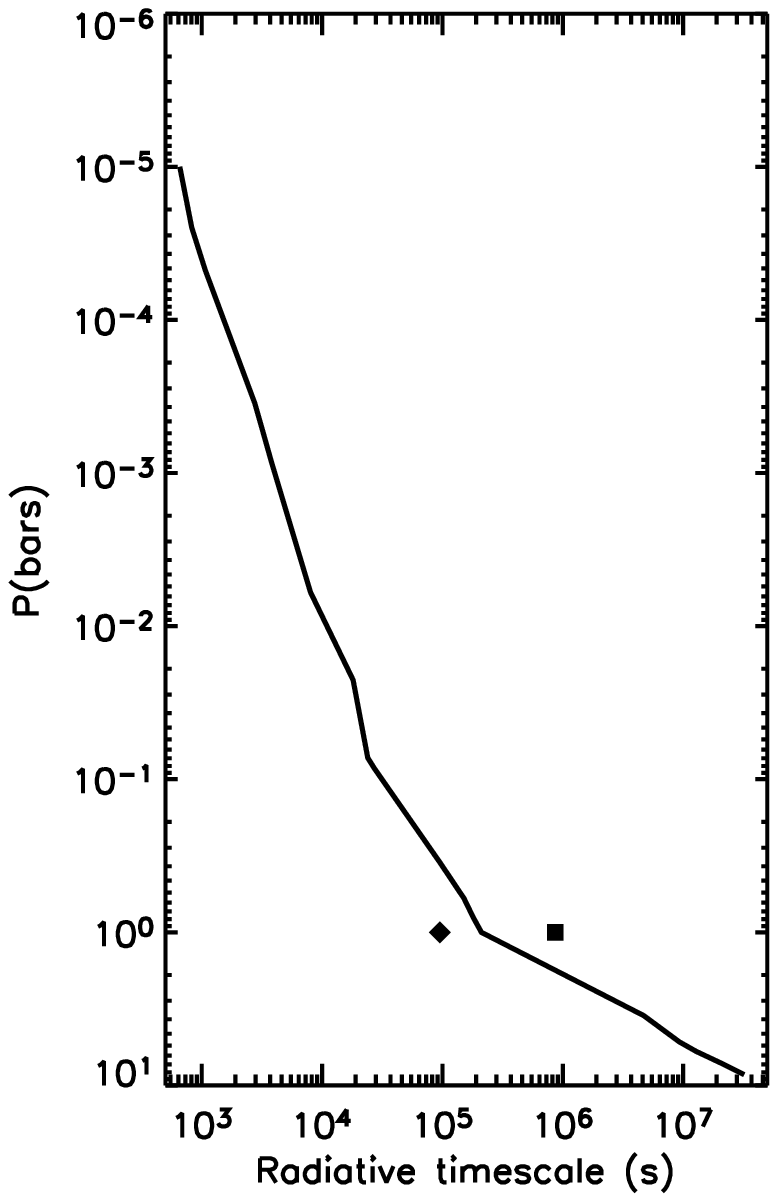}
\caption{Radiative timescale as a function of pressure
level\label{fig:tpsrad}.  The symbols show the values of the radiative
timescale at 1 bar from \citet{ShowGuillot} (1.1 day; diamond) and that of
\citet{Cho} (10 days; square).}
\end{center}
\end{figure}

The result is shown in Fig. \ref{fig:tpsrad}.  As expected, the
radiative time constant increases monotically with pressure. Around 1
bar, it is $\sim$2.3 days, 
about twice the value estimated by \citet{ShowGuillot}.
On the other hand, the long radiative timescale (10 days) assumed by \citet{Cho}
implies that they significantly underestimated the cooling in their
circulation model.
  
Above the 1--bar pressure level, the radiative
timescale is relatively short -- less than the rotation period -- and
we thus expect a rapid response from the planet's atmosphere to a
possible atmospheric circulation and large horizontal variations of
temperature. In particular, at 0.1 bar, $\tau_{\rm rad}$ is
3$\times$10$^4$ sec (8 hr). This region, which corresponds to the
optical limb of the planet at 0.6 $\mu$m due to Rayleigh scattering,
is that probed by the transit observations \citep{Charbonneau}. The
small radiative time constant suggests that the day--night thermal
contrast is large and that the atmospheric morning limb may be
significantly colder than the evening one for any reasonable zonal
wind speed -- as measured in the synchronously--rotating frame
--. This likely asymmetry, in temperature and hence potentially in the
chemical composition, should be kept in mind when analyzing
spectroscopic transits.  Below the 1--bar level, the opacity is large
and the timescale increases rapidly, varying almost as $p^2$. We
limited this calculation to 10 bar because below this level our
profile becomes convective in the nominal case (Fig.~\ref{fig:profs})
and the \emph{radiative} timescale then cannot be calculated with the
same method.

\section{Time-dependent solutions for an atmosphere in solid rotation}
\label{sec:timedep}

\subsection{Longitude-dependent temperature profiles}

\begin{figure}[p!]    
\begin{center}
   \includegraphics[width=8.8cm]{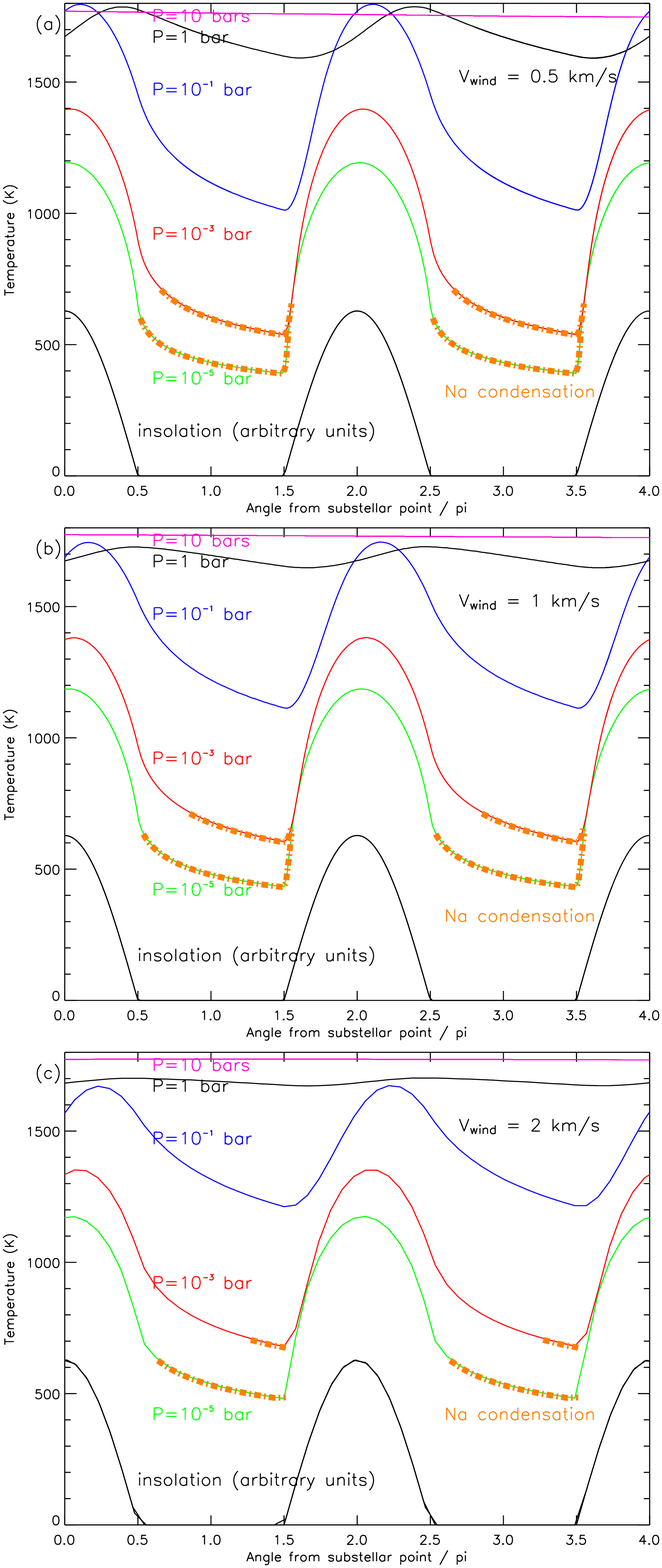}
\caption{Temperature at selected levels as a function of longitude for 
an equatorial wind velocity of (a) 0.5 km sec$^{-1}$ (b) 1 km sec$^{-1}$ 
and (c) 2 km sec$^{-1}$.  The incoming insolation is shown in 
arbitrary units. Note the shift of the maximum of temperature 
with respect to the maximum of insolation (at the substellar point) 
as depth (and so the radiative timescale) increases.  Below the 10--bar 
level, the temperature field is essentially uniform.\label{fig:rot}}
\end{center}
\end{figure}

We then introduce a solid body rotation by moving the atmosphere with 
a constant angular velocity with respect to the synchronously--rotating 
frame.  This procedure mimicks a zonal atmospheric 
circulation and allows us to investigate the temperature as a function 
of longitude. The effect of rotation 
is considered only through the modulation of the
incoming stellar flux.  The insolation is maximum at the substellar
longitude (noon) and constantly decreases until the atmosphere 
no longer receives the stellar flux on the night side. This periodic 
insolation is shown in Fig.~\ref{fig:rot}.  A first limitation 
of our model is that heating and cooling rates are valid 
for layers which are static with respect to each other.  The 
rotation of the atmosphere is thus approximated by a solid body
rotation, i.e.\ with a zonal wind constant with height. A second
limitation is that these heating and cooling rates correspond to
a given chemical composition (associated with the nominal temperature
profile in Fig.~\ref{fig:profs}) that does not change during the
integration process.

\begin{figure}[p!]    
\begin{center}
   \includegraphics[width=8.8cm]{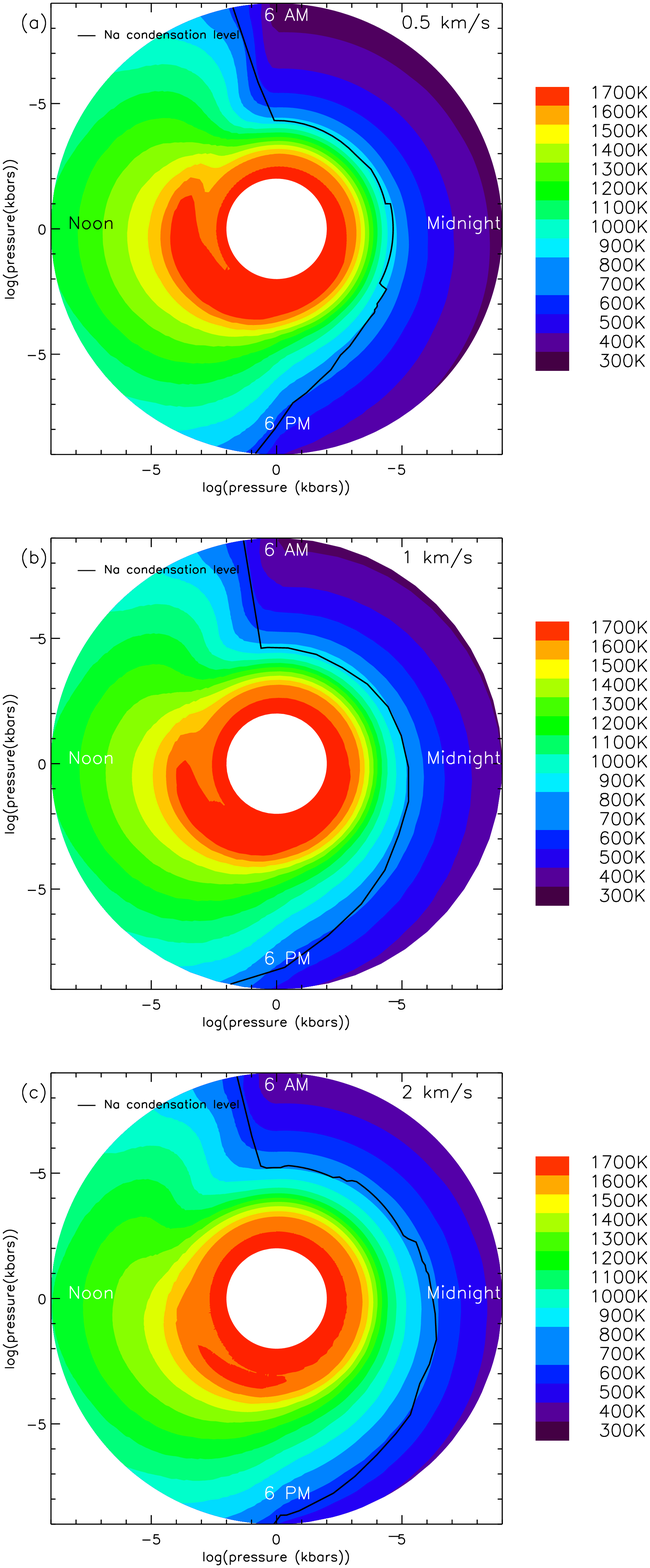}
\caption{Equatorial cut of the atmosphere between the 1$\times$10$^{-6}$ and
10--bar levels for an equatorial wind velocity of (a) 0.5 km sec$^{-1}$, 
(b) 1 km sec$^{-1}$, and (c) 2 km sec$^{-1}$. The level where 
condensation of sodium occurs (black line) goes deeper as the night
wears on (anti-clockwise) and is deepest at the morning limb 
\label{fig:rot2}. Below 10 bar, the temperature field (not shown here) 
is uniform and depends only on the bottom boundary condition.}
\end{center}
\end{figure}

\begin{figure}[tb]    
\begin{center}
   \includegraphics[width=\columnwidth]{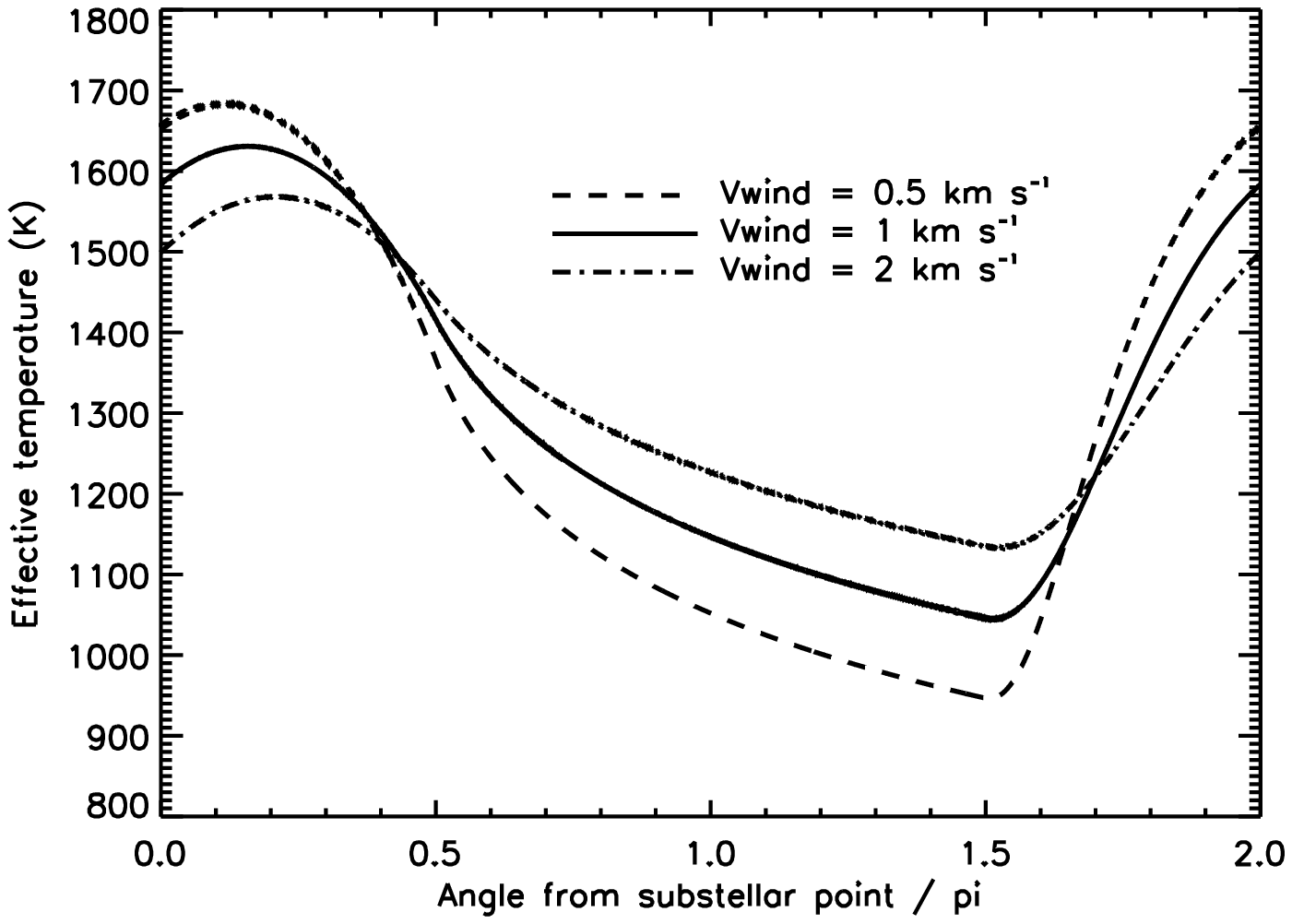}
\caption{Effective temperatures as a function of longitude for the three
 wind velocities.
\label{fig:teff}}
\end{center}
\end{figure}

We compute the temperature profiles for rotation periods corresponding 
to equatorial wind velocities of 0.5, 1 and 2 km sec$^{-1}$, in the
range predicted by \citet{ShowGuillot}. 
The effective temperature of the planet is shown in Fig.~\ref{fig:teff} 
and the temperature at selected levels is displayed in Fig.~\ref{fig:rot} 
as a function of longitude. 
Figure~\ref{fig:rot2} shows an equatorial cut of the atmosphere, with
the color scale for temperature indicated on the right.
The temperature maximum shifts with respect to the maximum 
of insolation as depth increases.  This phase lag results from 
the increase in the radiative timescale with depth.  It is 
particularly noticeable at $p \geq$~1 bar, where $\tau_{\rm rad}$ 
becomes comparable or exceeds the rotation period. We found 
day/night temperature variations exceeding  400, 600 and 800~K above the 
$\sim$0.1--bar level for equatorial winds of 2, 1 and 0.5 km sec$^{-1}$
respectively. These values are of the same order as predicted by 
\citet{ShowGuillot}.  However, at their reference level of 1 bar and 
for a wind of 1 km sec$^{-1}$, the contrast we get is at most 100~K, 
less than the $\sim$500~K they predict. Part of the discrepancy 
comes from the half lower radiative timescale used by these authors. 
In fact, the temperature contrast estimated by those authors pertains
to the level where $T \sim T_{\rm eff}$ which takes place at 0.15 bar
in our model, instead of 1 bar. Figure~5 shows that we do get contrasts
of 400--700~K in $T_{\rm eff}$ for winds in the range 2--0.5 km sec$^{-1}$.

Below the $\sim$5--bar level, the temperature field is essentially 
uniform with longitude (and time) due to the long radiative timescale. 
It is then equal to that given by the radiative equilibrium solution
under planet--averaged insolation conditions. This points to the need 
of considering this temperature as a boundary condition for evolutionary
models rather than that corresponding to the illuminated hemisphere
with no day--to--night redistribution of heat
(e.g., \citealp{Burrows03,Baraffe}).

An important effect is that the temperature profile becomes cold 
enough on the night side to allow sodium to condense, as shown
in Figs.~\ref{fig:rot}--\ref{fig:rot2}. The condensation level
extends relatively deep in the atmosphere on the night side and 
on the morning limb, down to 0.1, 0.3 and 0.5 bar for equatorial wind 
speeds of  2, 1 and 0.5 km sec$^{-1}$ respectively. This is close
or even below the optical limb around the Na resonance doublet at 589.3 nm.
Hence, during a transit, we expect stellar radiation at this wavelength 
to be less absorbed on the morning limb than on the evening limb.

\subsection{Sodium condensation and transit spectra}

\cite{Charbonneau} conducted HST
spectroscopic observations of \HD\ centered on the sodium resonance
doublet at 589.3 nm.  Using four planetary transits, they found a
visible deeper dimming in a bandpass centered on this feature than in
adjacent bands.  They interpreted it as absorption from sodium in the
planet's atmosphere.  However, the sodium absorption seen in the data
is lower than predicted by existing models by approximately 
a factor three \citep{Hubbard01,Brown01}.
These authors invoke several possibilities to explain such a depletion:

\begin{itemize}
\item condensation of sodium into mostly Na$_2$S. However according
to their model a very large fraction (99\%) of the sodium should 
condense out to explain by itself such a small absorption.
\item Ionization of sodium by the large stellar flux incident on the
planetary atmosphere.  This is a secondary effect which cannot explain
alone the weak absorption, as confirmed by \citet{Fortney03}.
\item Very high (above the 0.4-mbar level) particulate opacity.  This
possibility is difficult to assess as modeling of photochemical
hazes still needs to be performed.
\item A bulk depletion of the sodium abundance in \HD . It appears 
however that parent stars of close-in planets generally have a 
high metallicity \citep{Gonzalez97,Gonzalez00}.
\end{itemize}

An alternate explanation is provided by \citet{nonLTE}. According to
their work, Na could be far from being in local thermodynamic
equilibrium in the limb region. This effect would then reduce the
depth of the Na absorption seen during a transit. The amplitude of the
effect is however difficult to assess due to the lack of well
determined collisional deactivation rates from H$_2$ and He.

It should be noted that the models used to analyze the
observations were static and horizontally uniform whereas 
\citet{Charbonneau}'s data pertain to the planetary limb. Horizontal 
temperature variations due to insolation and dynamics can drive 
compositional variations that should be taken into account 
to calculate the sodium abundance over the limb. More specifically,
if sodium condenses on the night side, half of the limb is depleted 
in sodium above some level in the atmosphere.  This effect can
potentially reduce by half the sodium absorption during a transit 
compared to the prediction from static models.

\begin{figure}[tb]   
\begin{center}
   \includegraphics[width=\columnwidth]{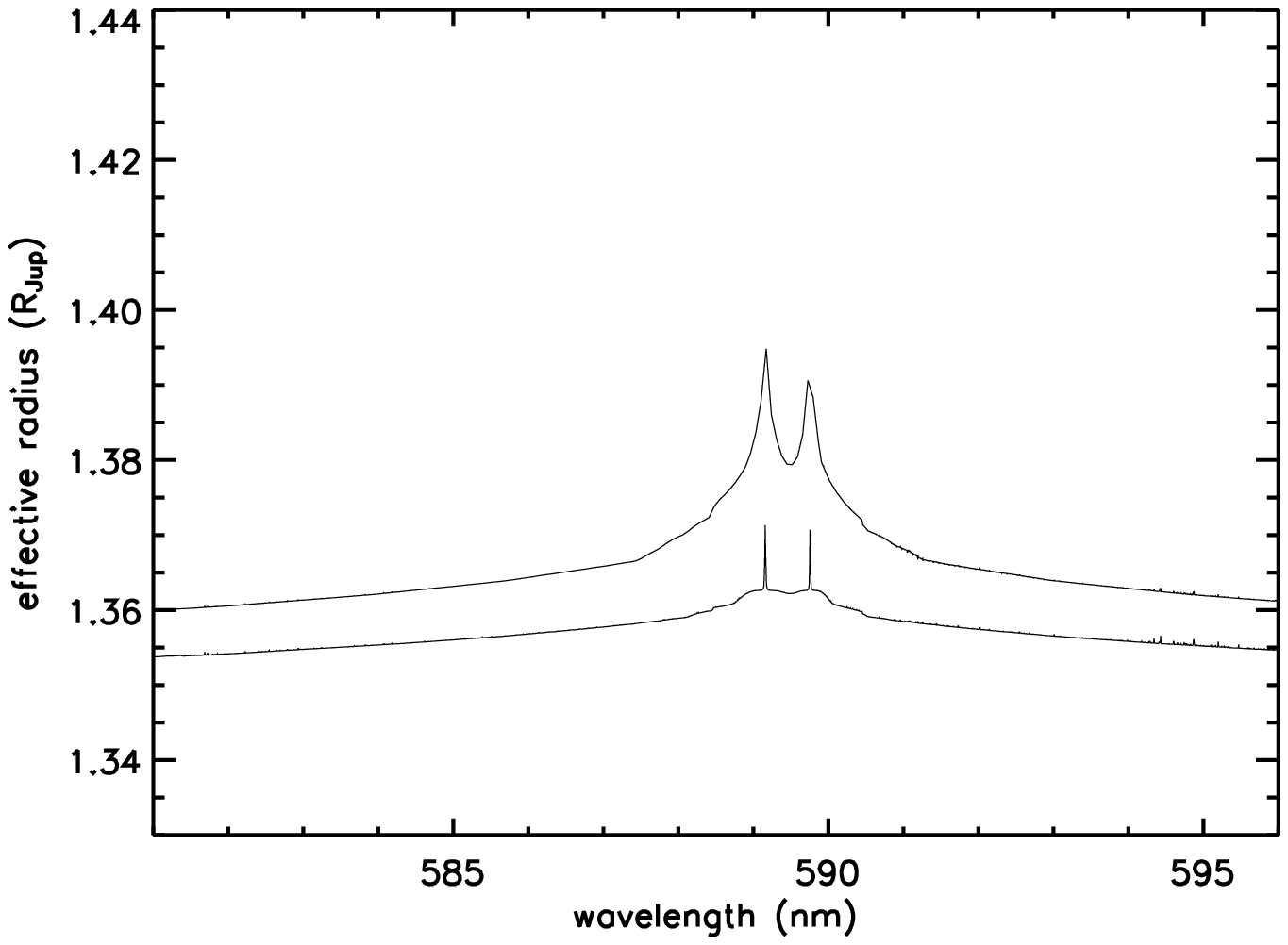}
\caption{\label{fig:Na}Spectra of the Na lines region for the morning limb (lower) and the evening limb (upper) for a wind velocity of 1 km/s.}
\end{center}
\end{figure}

We present here calculations of the spectral variation of the 
planetary radius in the region of the Na lines. 
The radius $r(\nu)$ is defined here as the limb 
corresponding to a tangential extinction optical depth of unity. 
To determine it, we calculate, as a function of frequency,
the transmission of the atmosphere for a series of lines of sight at
radii $r_{\rm i}(\nu)$ and then interpolate in this grid to find 
$r(\nu)$. This spectrum $r(\nu)$ is finally convolved to a resolution
of 1000. Two models are considered: one, corresponding 
to the evening limb, incorporates the temperature calculated at 
a phase of $\pi$/2 (see Fig.~\ref{fig:rot}) and the associated gas 
profiles; the second one corresponds to the morning limb at a phase 
of 3$\pi$/2. Results are shown in Fig.~\ref{fig:Na} for an equatorial wind
speed of 1 km sec$^{-1}$. As we had anticipated, the Na absorption 
is much less pronounced in the morning spectrum than in the evening 
one due to sodium condensation. Integrating the spectra over narrow 
bands at (588.7--589.9 nm) and around (581.8--588.7 nm and
589.9--596.8 nm) the Na feature as did \citet{Charbonneau}, we find 
that the absorption depth is three times weaker on the morning limb
($\delta R_{\rm m}=4.4 \times 10^{-3} R$) compared with the evening limb 
($\delta R_{\rm e}= 1.4 \times 10^{-2} R$).
During the transit, the stellar spectrum
is absorbed by the whole limb (morning and evening) and our model
thus predicts a dimming in the sodium band almost half lower than without
consensation of Na. We calculate it as $(\delta R_{\rm m} +
\delta R_{\rm e}) R /R_\star^2)$~=~2.8$\times$10$^{-4}$ within the uncertainty range of the observed value (2.32$\pm$0.57$\times$10$^{-4}$).

\section{Conclusions}
\label{conclusions}

We have developped a time-dependent one-dimensional radiative model and
applied it to the case of HD209458b. 
Because it is
one-dimensional, the model is necessarily simplified in the sense that
advection is included only horizontally in the limit of a solid-body
rotation of the atmosphere and vertically with convective adjustements
of the temperature profile where applicable. Furthermore, important
effects such as the possible presence of clouds and time-dependent
variations of the absorption coefficients due to variations of the
chemical composition are {\it not} included. These effects can be
considered in a useful way only through a detailed global circulation
model, which is beyond the scope of this paper. Conversely, the
results of our model can be used as a basis for future dynamical
models of the atmosphere of HD209458b and of Pegasides. They also
impact evolution calculations and our perception of the vertical and
longitudinal structure of strongly irradiated planets, as detailed
hereafter.

First, we point out that when considering constant 
planetary-averaged conditions, our radiative equilibrium profile
is colder than those calculated by \citet{Sudarsky03} and \citet{Baraffe}
mostly because those authors used a twice larger insolation. After correcting
to first order for this difference, our profile agrees with that of 
\citet{Sudarsky03} above the $\sim$0.3--mbar region, is still colder 
than \citet{Sudarsky03}'s and \citet{Baraffe}'s profile down to $\sim$0.3
bar (with a maximum discrepancy of 300--400~K), and is intermediate between
these two other models below this region. These remaining differences 
probably lie in the opacity sources or the radiative transfer treatment
but they should not alter the results that follow.

In our model, 99.99\% of the stellar flux is absorbed at the 
5--bar level (99 \% at the 2--bar level). This rapid absorption 
prevents a significant fraction ($\sim$1~\%) of the stellar energy 
from directly reaching pressure levels of several tens of bars. 
This mechanism cannot thus supply the energy source required 
at deep levels to explain the radius of \HD\ \citep{GuillotShow, Baraffe}.

The radiative timescales that we derive are generally twice longer
than those used by \citet{ShowGuillot} and 5 times shorter than assumed by \citet{Cho}.
Above the 1--bar level, the radiative time
constant is shorter than the rotation period of the planet.
These relatively short values imply that the atmosphere reacts 
relatively quickly to perturbations from atmospheric dynamics. 

In order to qualitatively test the effect of dynamics on the
atmospheric structure, we calculated models in which the stellar flux
was modulated with a period between 3.5 and 14 days, mimicking the effect of a
0.5-- to 2--km~s$^{-1}$ equatorial zonal jet on an otherwise synchronously
rotating planet.

 Depending on the imposed wind, we found longitudinal
temperature variations to be between 400 and 600~K at 0.1 bar, 30--200~K
at 1 bar and less than 5~K at 10 bar. This is generally consistent
with the results obtained by \cite{ShowGuillot}. On the other hand,
the fact that \cite{Cho} obtain temperatures on the night side that
can be hotter than on the day side is difficult to explain in light
of our results. This cannot be ruled out, but would presumably imply a
combination of strong meridional and vertical circulation.

The fact that the temperature rapidly becomes uniform with increasing
depth implies that the mixing most probably takes place in a
relatively shallow layer of the atmosphere. In our simulations, the
temperature reached at deep levels is consistent with that of an
atmosphere receiving a stellar flux averaged over the whole
planet on both the day side and the night side (i.e.\ 1/4 of
the stellar constant at the planet). This is very important for
evolution models, as temperature inhomogeneities at deep levels would
tend to fasten the cooling compared to homogeneous models with the
same absorbed luminosity \citep{GuillotShow}. On the contrary, some
evolution models \citep{Baraffe, Burrows03} are calculated from 
atmospheric boundary conditions that are relevant to the day side 
only. These models probably overestimate the temperature of the deep 
atmosphere and therefore the planet's cooling time. 

Finally, the large longitudinal temperature contrast implies that
species such as sodium will condense on the night side. Even with no
settling, the morning limb (which is coldest) can be strongly depleted
in the condensible species. From calculations of transit spectra representative 
of the morning and evening limbs, we found that the former shows a 3 times 
weaker sodium absorption than the latter. The Na dimming we calculated
through the entire limb is then in agreement with the sodium absorption 
observed by \citet{Charbonneau} during planetary transits.

\begin{acknowledgements}
This work was supported by the French Programme National de
Plan\'etologie of the Institut National des sciences de l'Univers (INSU).
\end{acknowledgements}

\bibliography{article.bbl}

\begin{thebibliography}{35}
\expandafter\ifx\csname natexlab\endcsname\relax\def\natexlab#1{#1}\fi

\bibitem[{{Anders} \& {Grevesse}(1989)}]{Anders}
{Anders}, E. \& {Grevesse}, N. 1989, \gca, 53, 197

\bibitem[Allard et al.(2000)]{Allard00} Allard, F., Hauschildt, 
P.~H., \& Schwenke, D.\ 2000, \apj, 540, 1005 

\bibitem[{{Baraffe} {et~al.}(2003){Baraffe}, {Chabrier}, {Barman}, {Allard}, \&
  {Hauschildt}}]{Baraffe}
{Baraffe}, I., {Chabrier}, G., {Barman}, T.~S., {Allard}, F., \& {Hauschildt},
  P.~H. 2003, \aap, 402, 701

\bibitem[{{Barman} {et~al.}(2001){Barman}, {Hauschildt}, \&
  {Allard}}]{Barman01}
{Barman}, T.~S., {Hauschildt}, P.~H., \& {Allard}, F. 2001, \apj, 556, 885

\bibitem[{{Barman} {et~al.}(2002){Barman}, {Hauschildt}, {Schweitzer},
  {Stancil}, {Baron}, \& {Allard}}]{nonLTE}
{Barman}, T.~S., {Hauschildt}, P.~H., {Schweitzer}, A., {et~al.} 2002, \apjl,
  569, L51

\bibitem[{{Bell}(1980)}]{Bell80}
{Bell}, K.~L. 1980, J. Phys. B: Atom. Molec. Phys., 13, 1859

\bibitem[{{Bodenheimer} {et~al.}(2003){Bodenheimer}, {Laughlin}, \&
  {Lin}}]{Boden03}
{Bodenheimer}, P., {Laughlin}, G., \& {Lin}, D.~N.~C. 2003, \apj, 592, 555

\bibitem[{{Bodenheimer} {et~al.}(2001){Bodenheimer}, {Lin}, \&
  {Mardling}}]{Boden01}
{Bodenheimer}, P., {Lin}, D.~N.~C., \& {Mardling}, R.~A. 2001, \apj, 548, 466

\bibitem[{{Brett}(1990)}]{Brett90}
{Brett}, J.~M. 1990, A\&A, 231, 440

\bibitem[{{Brown}(2001)}]{Brown01}
{Brown}, T.~M. 2001, \apj, 553, 1006

\bibitem[{{Brown} {et~al.}(2001){Brown}, {Charbonneau}, {Gilliland}, {Noyes},
  \& {Burrows}}]{Brown_al01}
{Brown}, T.~M., {Charbonneau}, D., {Gilliland}, R.~L., {Noyes}, R.~W., \&
  {Burrows}, A. 2001, \apj, 552, 699

\bibitem[{{Burrows} {et~al.}(1997){Burrows}, {Marley}, {Hubbard}, {Lunine},
  {Guillot}, {Saumon}, {Freedman}, {Sudarsky}, \& {Sharp}}]{Burrows97}
{Burrows}, A., {Marley}, M., {Hubbard}, W.~B., {et~al.} 1997, \apj, 491, 856

\bibitem[{{Burrows} {et~al.}(2000){Burrows}, {Marley}, \&
  {Sharp}}]{Burrows_alc}
{Burrows}, A., {Marley}, M.~S., \& {Sharp}, C.~M. 2000, ApJ, 531, 438

\bibitem[{{Burrows} {et~al.}(2003){Burrows}, {Sudarsky}, \&
  {Hubbard}}]{Burrows03}
{Burrows}, A., {Sudarsky}, D., \& {Hubbard}, W.~B. 2003, \apj, 594, 545

\bibitem[{{Charbonneau} {et~al.}(2000){Charbonneau}, {Brown}, {Latham}, \&
  {Mayor}}]{Charbonneau00}
{Charbonneau}, D., {Brown}, T.~M., {Latham}, D.~W., \& {Mayor}, M. 2000, ApJ,
  529, L45

\bibitem[{{Charbonneau} {et~al.}(2002){Charbonneau}, {Brown}, {Noyes}, \&
  {Gilliland}}]{Charbonneau}
{Charbonneau}, D., {Brown}, T.~M., {Noyes}, R.~W., \& {Gilliland}, R.~L. 2002,
  \apj, 568, 377

\bibitem[{{Cho} {et~al.}(2003){Cho}, {Menou}, {Hansen}, \& {Seager}}]{Cho}
{Cho}, J.~Y.-K., {Menou}, K., {Hansen}, B.~M.~S., \& {Seager}, S. 2003, \apjl,
  587, L117

\bibitem[{{Cody} \& {Sasselov}(2002)}]{CodySass02}
{Cody}, A.~M. \& {Sasselov}, D.~D. 2002, \apj, 569, 451

\bibitem[Fegley \& Lodders(1994)]{Fegley} Fegley, B.~J., \& 
Lodders, K.\ 1994, Icarus, 110, 117 

\bibitem[{{Fortney} {et~al.}(2003){Fortney}, {Sudarsky}, {Hubeny}, {Cooper},
  {Hubbard}, {Burrows}, \& {Lunine}}]{Fortney03}
{Fortney}, J.~J., {Sudarsky}, D., {Hubeny}, I., {et~al.} 2003, \apj, 589, 615

\bibitem[{{Gonzalez}(1997)}]{Gonzalez97}
{Gonzalez}, G. 1997, \mnras, 285, 403

\bibitem[{{Gonzalez}(2000)}]{Gonzalez00}
{Gonzalez}, G. 2000, in ASP Conf. Ser. 219: Disks, Planetesimals, and Planets,
  523

\bibitem[{{Goukenleuque} {et~al.}(2000){Goukenleuque}, {B{\' e}zard}, {Joguet},
  {Lellouch}, \& {Freedman}}]{Gouken}
{Goukenleuque}, C., {B{\' e}zard}, B., {Joguet}, B., {Lellouch}, E., \&
  {Freedman}, R. 2000, Icarus, 143, 308

\bibitem[{{Guillot} {et~al.}(1996){Guillot}, {Burrows}, {Hubbard}, {Lunine}, \&
  {Saumon}}]{Guillot96}
{Guillot}, T., {Burrows}, A., {Hubbard}, W.~B., {Lunine}, J.~I., \& {Saumon},
  D. 1996, \apjl, 459, L35

\bibitem[{{Guillot} {et~al.}(1994){Guillot}, {Gautier}, {Chabrier}, \&
  {Mosser}}]{Guillot94}
{Guillot}, T., {Gautier}, D., {Chabrier}, G., \& {Mosser}, B. 1994, Icarus,
  112, 337

\bibitem[{{Guillot} \& {Showman}(2002)}]{GuillotShow}
{Guillot}, T. \& {Showman}, A.~P. 2002, \aap, 385, 156

\bibitem[{{Henry} {et~al.}(2000){Henry}, {Marcy}, {Butler}, \&
  {Vogt}}]{Henry00}
{Henry}, G.~W., {Marcy}, G.~W., {Butler}, R.~P., \& {Vogt}, S.~S. 2000, ApJ,
  529, L41

\bibitem[{{Hubbard} {et~al.}(2001){Hubbard}, {Fortney}, {Lunine}, {Burrows},
  {Sudarsky}, \& {Pinto}}]{Hubbard01}
{Hubbard}, W.~B., {Fortney}, J.~J., {Lunine}, J.~I., {et~al.} 2001, \apj, 560,
  413

\bibitem[{{John}(1988)}]{John88}
{John}, T.~L. 1988, A\&A, 193, 189

\bibitem[{{Kurucz}(1970)}]{Kurucz}
{Kurucz}, R.~L. 1970, SAO Special Report, 308

\bibitem[Lodders(1999)]{Lodders} Lodders, K.\ 1999, \apj, 519, 
793 

\bibitem[{{Mazeh} {et~al.}(2000){Mazeh}, {Naef}, {Torres}, {Latham}, {Mayor},
  {Beuzit}, {Brown}, {Buchhave}, {Burnet}, {Carney}, {Charbonneau}, {Drukier},
  {Laird}, {Pepe}, {Perrier}, {Queloz}, {Santos}, {Sivan}, {Udry}, \&
  {Zucker}}]{Mazeh00}
{Mazeh}, T., {Naef}, D., {Torres}, G., {et~al.} 2000, ApJ, 532, L55

\bibitem[{{Pierce} \& {Allen}(1977)}]{PierceAllen}
{Pierce}, A.~K. \& {Allen}, R.~G. 1977, in The Solar Output and its Variation,
  169--192

\bibitem[{{Robichon} \& {Arenou}(2000)}]{Rob_Arenou}
{Robichon}, N. \& {Arenou}, F. 2000, \aap, 355, 295

\bibitem[{{Seager} \& {Sasselov}(1998)}]{Seag_Sass98}
{Seager}, S. \& {Sasselov}, D.~D. 1998, \apjl, 502, L157

\bibitem[{{Seager} \& {Sasselov}(2000)}]{Seag_Sass00}
{Seager}, S. \& {Sasselov}, D.~D. 2000, \apj, 537, 916

\bibitem[{{Showman} \& {Guillot}(2002)}]{ShowGuillot}
{Showman}, A.~P. \& {Guillot}, T. 2002, A\&A, 385, 166

\bibitem[{{Sudarsky} {et~al.}(2003){Sudarsky}, {Burrows}, \&
  {Hubeny}}]{Sudarsky03}
{Sudarsky}, D., {Burrows}, A., \& {Hubeny}, I. 2003, \apj, 588, 1121

\end{thebibliography}
\end{document}